\documentstyle[aps,12pt,axodraw]{revtex}

\catcode`@=12
\newcommand{\beq}{\begin{equation}}
\newcommand{\eeq}{\end{equation}}

\newcommand{\bea}{\begin{eqnarray}}
\newcommand{\eea}{\end{eqnarray}}

\newcommand{\np}[1]{{ Nucl. Phys. }{\bf #1}}

\begin{document}
\thispagestyle{empty}
\begin{flushright} DESY 98-166\\ UCRHEP-T242\\January 1999\
\end{flushright}
\vspace{0.5in}
\begin{center}
{\Large \bf Neutrino Masses in Supersymmetry:\\R-Parity and Leptogenesis\\}
\vspace{1.0in}
{\bf Ernest Ma$^1$, Martti Raidal$^2$, and Utpal Sarkar$^3$\\}
\vspace{0.2in}
{$^1$ \sl Department of Physics, University of California\\}
{\sl Riverside, California 92521, USA\\}
\vspace{0.1in}
{$^2$ \sl Theory Group, DESY, D-22603 Hamburg, Germany\\}
\vspace{0.1in}
{$^3$ \sl Physical Research Laboratory, Ahmedabad 380 009, India\\}
\vspace{1.0in}
\end{center}
\begin{abstract}
In the supersymmetric standard model of particle interactions, R-parity 
nonconservation is often invoked to obtain nonzero neutrino masses.  We point 
out here that such interactions of the supersymmetric particles would erase 
any pre-existing lepton or baryon asymmetry of the universe before the 
electroweak phase transition through the $B + L$ violating sphaleron 
processes.  We then show how neutrino masses may be obtained in supersymmetry 
(assuming R-parity conservation) together with successful leptogenesis and 
predict the possible existence of new observable particles.
\end{abstract}

\newpage
\baselineskip 24pt

Two issues in particle physics are critically important today.  One is the 
possible existence of neutrino masses, as evidenced by the ongoing 
excitement generated by the recent report of atmospheric neutrino 
oscillations \cite{1}, as well as previous other indications of solar \cite{2} 
and accelerator \cite{3} neutrino oscillations.  The other is the possible 
existence of supersymmetry, as evidenced by the enormous, continuing efforts 
of both experimentalists and theorists in devising ways of searching for the 
predicted new particles in existing and future high-energy colliders \cite{4}. 
In the minimal standard model (SM) 
of quarks and leptons without supersymmetry, 
neutrinos are massless.  To make them massive, new physics have to be 
assumed \cite{5}.  In the minimal supersymmetric standard model (MSSM) which 
assumes R-parity conservation, neutrinos are also massless.  To make them 
massive, many previous discussions have been based on R-parity 
nonconservation \cite{6}.  We point out here one very important consequence 
of this hypothesis, namely that there are now unavoidable lepton-number 
violating interactions at the supersymmetry breaking scale.  Combining these 
with the $B + L$ violating sphaleron processes \cite{7}, any pre-existing 
$B$ or $L$ or $B - L$ asymmetry of the universe would be erased \cite{7a,8}.  
This is so unless $B - 3L_i$ is conserved \cite{9,10} even after the 
electroweak phase transition, which is of course not the case here.  
A more desirable mechanism for neutrino masses in supersymmetry should be 
such that leptogenesis \cite{11} would be also possible in the same context.

There are two appealing mechanisms for neutrino masses which are 
intimately related to successful leptogenesis.   One is the canonical 
seesaw mechanism \cite{12}, in which heavy Majorana singlet neutrinos may 
decay to generate a lepton asymmetry \cite{13}, which gets converted into 
the present observed baryon asymmetry through the electroweak phase 
transition.  The other is to have neutrino 
masses as well as leptogenesis through heavy Higgs triplets \cite{14}. 
Both are conceived originally as simple extensions of the SM, 
but they are also applicable as simple extensions of the MSSM.  The key 
in both cases is that lepton-number violation occurs at a mass scale many 
orders of magnitude greater than the electroweak breaking scale of $10^2$ GeV. 
In models of R-parity violation, the participating particles are those of 
the MSSM, hence lepton-number violation is a fast process at the supersymmetry 
breaking scale of $10^3$ GeV.  We show in the following that for realistic 
neutrino masses, such models do not allow leptogenesis.  Included in this 
class of models are those which obtain neutrino masses from a 
radiative mechanism with suppressed Yukawa couplings, such as the Zee model 
\cite{15,15a,15b}.  We then propose a specific supersymmetric extension of the 
Zee model with \underline {unsuppressed} Yukawa couplings which has new 
particles at a much higher mass scale.  We demonstrate the possibility of 
obtaining realistic radiative neutrino masses as well as successful 
leptogenesis in this model.  It also contains other new particles that 
should be light enough for discovery at future accelerators such as the 
planned Large Hadron Collider (LHC) at CERN.

In the MSSM, R-parity of a particle is defined as
\begin{equation}
R \equiv (-1)^{3B + L + 2J},
\end{equation}
where $B$ is its baryon number, $L$ its lepton number, and $J$ its 
spin angular momentum.  Hence the SM particles have 
$R = +1$ and their supersymmetric partners have $R = -1$.  Using the 
common notation where all chiral superfields are considered left-handed, 
the three families of leptons and quarks are given by
\begin{equation}
L_i = (\nu_i,e_i) \sim (1,2,-1/2), ~~~ e^c_i \sim (1,1,1),
\end{equation}
\begin{equation}
Q_i = (u_i,d_i) \sim (3,2,1/6), ~~~ u^c_i \sim (3^*,1,-2/3), ~~~ d^c_i \sim 
(3^*,1,1/3),
\end{equation}
where $i$ is the family index, and the two Higgs doublets are given by
\begin{equation}
H_1 = (h^0_1,h^-_1) \sim (1,2,-1/2), ~~~ H_2 = (h^+_2,h^0_2) \sim (1,2,1/2),
\end{equation}
where the $SU(3)_C \times SU(2)_L \times U(1)_Y$ content of each superfield 
is also indicated.  If R-parity is conserved, the superpotential is 
restricted to have only the terms
\begin{equation}
W = \mu H_1 H_2 + f^e_{ij} H_1 L_i e^c_j + f^d_{ij} H_1 Q_i d^c_j + f^u_{ij} 
H_2 Q_i u^c_j.
\end{equation}
If R-parity is violated but not baryon number, then the superpotential 
contains the additional terms
\begin{equation}
W' = \epsilon_i L_i H_2 + \lambda_{ijk} L_i L_j e_k^c + \lambda'_{ijk} L_i Q_j 
d_k^c,
\end{equation}
resulting in nonzero neutrino masses either from mixing with the neutralino 
mass matrix \cite{6} or in one-loop order \cite{16}.

If lepton-number violating interactions such as
\begin{equation}
L_i + Q_j \to (\tilde d^c_k)^* \to H_1 + Q_l
\end{equation}
are in equilibrium in the early universe, any pre-existing lepton asymmetry 
would be erased.  To make sure that this does not happen, the following 
condition has to be satisfied:
\begin{equation}
{\lambda'^2 T \over 8 \pi} \lesssim 1.7 \sqrt{g_*} {T^2 \over M_P} ~~~ {\rm 
at}~~T = M_{SUSY},
\end{equation}
where $g_* \sim 10^2$ is the effective number of interacting relativistic 
degrees of freedom and $M_P \sim 10^{19}$ GeV is the Planck mass.  Assuming 
that the supersymmetry breaking scale $M_{SUSY}$ is $10^3$ GeV, we find
\begin{equation}
\lambda' \lesssim 2 \times 10^{-7},
\end{equation}
which is very much below the typical minimum value of $10^{-4}$ needed for 
radiative neutrino masses \cite{17}.  A similar bound was presented from
dimensional arguments \cite{7a}.  Larger values of $\lambda'$ are 
allowed if there is a conserved $(B-3L_i)$ symmetry \cite{9}.  However, 
there would be other severe phenomenological restrictions in that case 
\cite{17a}.  The bound of Eq.~(9) cannot be evaded even if one uses the 
bilinear term in Eq.~(6) for neutrino masses instead, because the induced 
mixing would change Eq.~(5) and introduce trilinear couplings which violate 
lepton number and an effective $\lambda'$ is unavoidable. 

Because the $B + L$ violating sphaleron processes are effective at 
temperatures from $10^2$ to $10^{12}$ GeV, the presence of the above $L$ 
violating interactions would also erase any pre-existing $B$ or $B - L$ 
asymmetry of the universe before the electroweak phase transition.  To have 
the successful conversion of a pre-existing $B$ or $L$ or $B - L$ asymmetry 
into the present observed baryon asymmetry of the universe, it is necessary 
that the lepton-number violating interactions of Eq.~(6) be suppressed.  This 
means that although R-parity violation may exist, it will mostly be negligible 
phenomenologically.  In particular, it will not contribute significantly 
to neutrino masses.

It is clear from the above discussion that we need to increase the mass scale 
of any appreciable lepton-number violating interaction for it to be consistent 
with leptogenesis.  Of course we would also like it to generate appropriate 
neutrino masses.  As remarked earlier, such models \cite{13,14} are already 
well-known.  Whether heavy Majorana singlet neutrinos or heavy Higgs triplets 
are used, the scale of lepton-number violation is determined by their masses, 
which may be greater than $10^7$ GeV or $10^{13}$ GeV in the case of the 
former or the latter respectively.  In both cases, there are no new observable 
particles or interactions below that scale.  Extending these models to include 
supersymmetry \cite{18} does not change the above conclusion, other than the 
obvious fact that the SM now becomes the MSSM.  Indeed it should be noted that 
R-parity is conserved in both such extensions, because lepton number is 
violated only by two units.  On the other hand, models of radiative neutrino 
masses often include new particles which are amenable to discovery at 
planned future accelerators.  They may also offer the possibility of 
naturally large mixing angles which are required for atmospheric 
neutrino oscillations and for vacuum solar neutrino oscillations. 

In models of radiative neutrino masses \cite{5,19}, in addition to the 
suppression due to the $1/16 \pi^2$ factor of each loop, there is often 
another source of suppression due to the Yukawa couplings involved. In the 
supersymmetric case with R-parity violation \cite{16}, the suppression is 
proportional to a quadratic combination of charged-lepton or down-quark 
masses \cite{17}.  That is the reason why $\lambda$ and $\lambda'$ of Eq.~(6) 
cannot be too small.  In the original Zee model \cite{15}, the SM is 
extended to include a charged scalar $\chi^+$ and a second Higgs doublet. 
The radiative mechanism for generating neutrino masses is exactly the same 
\cite{5} as given by Eq.~(6) with $e^c$ replaced by $\chi^+$. 
Although the mass of $\chi^+$ is not constrained by $M_{SUSY}$, the 
previously mentioned Yukawa suppression factor remains, hence it cannot 
be too large or else neutrino masses would be too small, as shown below.

The relevant terms of the interaction Lagrangian are given by
\begin{equation}
{\cal L} = \sum_{i < j} f_{ij} (\nu_i e_j - e_i \nu_j) \chi^+ + \mu 
(\phi_1^+ \phi_2^0 - \phi_1^0 \phi_2^+) \chi^- + H.c.,
\end{equation}
where two Higgs doublets are needed or else there would be no $\phi \phi 
\chi$ coupling.  Lepton number is violated in the above by two units, hence 
we expect the realization of an effective dimension-five operator 
$\Lambda^{-1} \phi^0 \phi^0 \nu_i \nu_j$ for naturally small Majorana 
neutrino masses \cite{5}.  This occurs here in one loop and the elements of 
the $3 \times 3$ neutrino mass matrix are given by
\begin{equation}
(m_\nu)_{ij} = f_{ij} (m_i^2 - m_j^2) \left( {\mu v_2 \over v_1} \right) 
F(m_\chi^2,m_{\phi_1}^2),
\end{equation}
where $v_{1,2} \equiv \langle \phi^0_{1,2} \rangle$ and $m_i$ are the 
charged-lepton masses which come from $\phi_1$ but not $\phi_2$.  The 
function $F$ is given by
\begin{equation}
F(m_1^2, m_2^2) = {1 \over 16 \pi^2} {1 \over m_1^2 - m_2^2} \ln {m_1^2 \over 
m_2^2}.
\end{equation}
Since the $m_\tau^2$ terms in Eq.~(11) are likely to be dominant, this model 
has two nearly mass-degenerate neutrinos which mix maximally \cite{15a,21}. 
This is very suitable for explaining the atmospheric neutrino data \cite{1}, 
but only in conjunction with the LSND data \cite{3}.  Let $m_\chi = 1$ TeV, 
$m_{\phi_1} = 100$ GeV, $\mu = 100$ GeV, $v_2/v_1 = 1$, and $f_{\mu \tau} = 
f_{e \tau} = 10^{-7}$ to satisfy Eq.~(9), then the $m_\tau^2$ terms 
generate a neutrino mass of 0.0013 eV, which is very much below the necessary 
1 eV or so indicated by the LSND data.  We note that Eq.~(8) constrains 
the combination $f^2/m_\chi$, whereas $m_\nu$ goes like $f/m_\chi^2$.  
Hence neutrino masses would only decrease if we increase $m_\chi$.  As long 
as Eq.~(11) gets a suppression from $m_\tau^2$ (which comes of course from 
the Yukawa coupling $m_\tau/v_1$), the conflict with leptogenesis is a real 
problem.

We now propose a new supersymmetric variation of the Zee model which has a 
fourth family of leptons with unsuppressed Yukawa couplings for generating 
neutrino masses (the fourth quark family should be also added to cancel 
anomalies but we do not consider their phenomenology here).  
This model preserves R-parity and the scale of lepton-number 
violation by two units is of order $10^{13}$ GeV, which is suitable for 
leptogenesis.  We add to the MSSM the new superfields shown in Table I. 
The discrete $Z_2$ symmetry is just the usual one for defining R-parity; 
{\it i.e.} the quark and lepton superfields are odd and the Higgs superfields 
are even.  The discrete $Z'_2$ symmetry is new and it distinguishes the new 
particles of Table I from those of the MSSM which are assumed to be even.

The relevant terms in the R-parity preserving superpotential of this model 
are given by
\begin{eqnarray}
W &=& \mu_{12} (h_1^0 h_2^0 - h_1^- h_2^+) + \mu_{34} (h_3^0 h_4^0 - h_3^- 
h_4^+) + m_\chi \chi_1^+ \chi_2^- + (m_E/v_1) (h_1^0 E^- - h_1^- N_1^0) E^+ 
\nonumber \\ && + f_i (\nu_i h_3^- - e_i^- h_3^0) E^+ + f'_j (\nu_j E^- - 
e_j^- N_1^0) \chi_1^+ + f_{24} (h_2^+ h_4^0 - h_2^0 h_4^+) \chi_2^-,
\label{spot}
\end{eqnarray}
and $v_{1,2}$ are the vacuum expectation values of $h^0_{1,2}$. 
The unsuppressed one-loop diagram generating neutrino masses is shown in 
Fig.~1.  We note that the effective dimension-five operator $L_i L_j H_2 H_2$ 
is indeed realized.  Assuming that the masses of the scalar leptons of the 
fourth family to be equal to $M_{SUSY}$, we find
\begin{equation}
(m_\nu)_{ij} = {(f_i f'_j + f'_i f_j) f_{24} v_2^2 m_E \mu_{12} \mu_{34} \over 
16 \pi^2 v_1 M_{SUSY}^2 m_\chi} \ln {m_\chi^2 \over M_{SUSY}^2}.
\end{equation}
To get an estimate of the above expression, let $f_i = f'_j = f_{24} = 1$, 
$m_E = v_1$, $\mu_{12} = \mu_{34} = M_{SUSY}$, then
\begin{equation}
m_\nu = {1 \over 8 \pi^2} {v_2^2 \over m_\chi} \ln {m_\chi^2 \over 
M_{SUSY}^2}.
\end{equation}
Assuming $v_2 \sim 10^2$ GeV, $m_\chi \sim 10^{13}$ GeV, and $M_{SUSY} \sim 
10^3$ GeV, we get $m_\nu \sim 0.6$ eV.  This is just one order of magnitude 
greater than the square root of the $\Delta m^2 \sim 5 \times 10^{-3}$ 
eV$^2$ needed for atmospheric neutrino oscillations \cite{1}.  Reducing 
slightly the above dimensionless couplings from unity would fit the data 
quite well.  Since $m_\chi \sim 10^{13}$ GeV is now allowed, leptogenesis 
should be possible as demonstrated in \cite{8}. It was argued 
\cite{gravitino} that due to the gravitino production constraints on the
reheating temperature after the inflation such a 
high leptogenesis scale is allowed only in models with small gravitino
masses, e.g., models with gauge mediated SUSY breaking. However, 
new efficient reheating mechanisms \cite{linde} allow production of
particles with such masses, and consequently leptogenesis, 
without exponential suppression.

It has recently been shown \cite{22} that the structure of Eq.~(14) for the 
$\mu - \tau$ sector is naturally suited for the large mixing solution of 
atmospheric neutrino oscillations.  To be more specific, the $2 \times 2$ 
submatrix of Eq.~(14) for the $\mu - \tau$ sector can be written as
\begin{equation}
{\cal M} = m_0 \left[ \begin{array} {c@{\quad}c} 2 \sin \alpha \sin \alpha' & 
\sin (\alpha + \alpha') \\ \sin (\alpha + \alpha') & 2 \cos \alpha \cos 
\alpha' \end{array} \right],
\end{equation}
where $\tan \alpha = f_\mu/f_\tau$ and $\tan \alpha' = f'_\mu/f'_\tau$. 
The eigenvalues of $\cal M$ are then given by $m_0 (c_1 \pm 1)$, where 
$c_1 = \cos (\alpha - \alpha')$, and the effective $\sin^2 2 \theta$ for 
$\nu_\mu - \nu_\tau$ oscillations is $(1-c_2)/(1+c_2)$, where 
$c_2 = \cos (\alpha + \alpha')$.  If we choose $\tan \alpha \sim \tan \alpha' 
\sim 1$, then $c_1 \sim 1$ and $c_2 \sim 0$.  In that case, maximal mixing 
between a heavy ($2 m_0$) and a light ($s_1^2 m_0/2$) neutrino occurs as 
an explanation of the atmospheric data.  If we assume further that $f_e << 
f_{\mu,\tau}$ and $f'_e << f'_{\mu,\tau}$, then the small-angle 
matter-enhanced solution of solar neutrino oscillations may be obtained as 
well.  

Our proposed model has the twin virtues of an acceptable neutrino mass matrix 
given by Eq.~(14) and the possibility of generating a lepton asymmetry of the 
universe through the decays of $\chi^\pm_{1,2}$.  It is also 
phenomenologically safe because all the additions to the SM do not alter its 
known successes.  Neither the fourth family of leptons $E^\pm$, $N^0_{1,2}$ 
nor the two extra Higgs doublets $H_{3,4}$ mix with their SM analogs because 
they are odd under the new discrete $Z'_2$ symmetry.  In particular, $H_3$ 
and $H_4$ do not couple to the known quarks and leptons, hence 
flavor-changing neutral currents are suppressed here as in the SM.  The 
lepton-number violation of this model is associated with $m_\chi$ which is 
of order $10^{13}$ GeV.  However, the fourth family of leptons should have 
masses of order $10^2$ GeV and be observable at planned future colliders. 
The two extra Higgs doublets should also be observable with an energy scale 
of order $M_{SUSY}$.  The soft supersymmetry-breaking terms of this model 
are assumed to break $Z'_2$ without breaking $Z_2$.  Hence there will 
still be a stable LSP (lightest supersymmetric particle) and a fourth-family 
lepton will still decay into ordinary leptons.  For example, because 
$\tilde h_3^0$ mixes with $\tilde h_1^0$, the decay
\begin{equation}
E^- \to \mu^- \tilde h_3^0 ~(\tilde h_1^0) \to \mu^- \tau^+ \tau^-
\end{equation}
is possible and would make a spectacular signature.

In conclusion, we have shown in this paper that R-parity violation in 
supersymmetry is generically inconsistent with leptogenesis because the 
lepton-number violating interactions would act in conjunction with the 
$B + L$ violating sphaleron processes and erase any pre-existing $B$ or $L$ 
or $B - L$ asymmetry of the universe.  This constraint means that any 
R-parity violation must be very small, so that it is of negligible 
phenomenological interest and cannot contribute significantly to neutrino 
masses.  This conclusion also applies to models of radiative neutrino masses 
with suppressed Yukawa couplings, such as the Zee model.  However, we have 
also shown that realistic radiative neutrino masses in supersymmetry are 
possible beyond the MSSM with R-parity conservation where the lepton-number 
violation is by two units and occurs at the mass scale of $10^{13}$ GeV.  
Our specific model (which is an extension of the Zee model with unsuppressed 
Yukawa couplings) also predicts new particles which should be observable in 
the future at the LHC.

\acknowledgements{ We thank W. Buchm\"uller for useful comments on 
the manuscript.
The work of EM was supported in part by the U.~S.~Department 
of Energy under Grant No.~DE-FG03-94ER40837, and MR acknowledges financial 
support from the Alexander von Humboldt Foundation.}

\bibliographystyle{unsrt}

\begin{thebibliography}{99}
\bibitem{1} Y. Fukuda {\it et al.}, Phys. Lett. {\bf B433}, 9 (1998); 
{\bf B436}, 33 (1998); Phys. Rev. Lett. {\bf 81}, 1562 (1998).
\bibitem{2} R. Davis, Prog. Part. Nucl. Phys. {\bf 32}, 13 (1994); Y. Fukuda 
{\it etal.}, Phys. Rev. Lett. {\bf 77}, 1683 (1996); {\bf 81}, 1158 (1998); 
P. Anselmann {\it etal.}, Phys. Lett. {\bf B357}, 237 (1995); {\bf B361}, 235 
(1996); J. N. Abdurashitov {\it et al.}, Phys. Lett. {\bf B328}, 234 (1994).
\bibitem{3} C. Athanassopoulos {\it et al.}, Phys. Rev. Lett. {\bf 75}, 2650 
(1995); {\bf 77}, 3082 (1996); {\bf 81}, 1774 (1998).
\bibitem{4} For a recent review and a list of references, see for example 
V. Barger, hep-ph/9801440.
\bibitem{5} For a recent overview, see E. Ma, Phys. Rev. Lett. {\bf 81}, 1171 
(1998).
\bibitem{6} For a recent review, see for example J. W. F. Valle, 
hep-ph/9808292.
\bibitem{7} V. A. Kuzmin, V. A. Rubakov, and M. E. Shaposhnikov, Phys. Lett. 
{\bf 155B}, 36 (1985).
\bibitem{7a} B. A. Campbell, S. Davidson, J. E. Ellis and K. Olive,
Phys. Lett. {\bf B256}, 457 (1991).
\bibitem{8} E. Ma, M. Raidal, and U. Sarkar, hep-ph/9811240.
\bibitem{9} H. Dreiner and G. G. Ross, Nucl. Phys. {\bf B410}, 188 (1993); 
A. Ilakovac and A. Pilaftsis, Nucl. Phys. {\bf B437}, 491 (1995).
\bibitem{10} E. Ma and U. Sarkar, Phys. Lett. {\bf B439}, 95 (1998).
\bibitem{11} For recent reviews, see for example, U. Sarkar, hep-ph/9809209;
W. Buchm\"uller, hep-ph/9812447, and references therein.
\bibitem{12} M. Gell-Mann, P. Ramond, and R. Slansky, in {\em Supergravity}, 
edited by P. van Nieuwenhuizen and D. Z. Freedman (North-Holland, Amsterdam, 
1979), p.~315; T. Yanagida, in {\em Proceedings of the Workshop on the Unified 
Theory and the Baryon Number in the Universe}, edited by O. Sawada and 
A. Sugamoto (KEK Report No.~79-18, Tsukuba, Japan, 1979), p.~95; R. N. 
Mohapatra and G. Senjanovic, Phys. Rev. Lett. {\bf 44}, 1316 (1980).
\bibitem{13} M. Fukugita and T. Yanagida, Phys. Lett. {\bf B174}, 45 (1986).
\bibitem{14} E. Ma and U. Sarkar, Phys. Rev. Lett. {\bf 80}, 5716 (1998).
\bibitem{15} A. Zee, Phys. Lett. {\bf 93B}, 389 (1980).
\bibitem{15a} A. Yu. Smirnov and M. Tanimoto, Phys. Rev. {\bf D55}, 1665 
(1997).
\bibitem{15b} C. Jarlskog, M. Matsuda, S. Skadhauge, and M. Tanimoto, 
hep-ph/9812282.
\bibitem{16} L. Hall and M. Suzuki, Nucl. Phys. {\bf B231}, 419 (1984).
\bibitem{17} For a recent discussion, see for example M. Drees, S. Pakvasa, 
X. Tata, and T. ter Veldhuis, Phys. Rev. {\bf D57}, R5340 (1998).
\bibitem{17a} E. Ma, Phys. Lett. {\bf B433}, 74 (1998); E. Ma and D. P. Roy, 
Phys. Rev. {\bf D58}, 095005 (1998)..
\bibitem{18} See for example, M. Pl\"umacher, \np{B530}, 207 (1998);
G. Lazarides and Q. Shafi, Phys. Rev. {\bf D58}, 
071702 (1998); Q. Shafi and Z. Tavartkiladze, hep-ph/9811463.
\bibitem{19} K. S. Babu and E. Ma, Mod. Phys. Lett. {\bf A4}, 1975 (1989).
\bibitem{21} N. Gaur, A. Ghosal, E. Ma, and P. Roy, Phys. Rev. {\bf D58}, 
071301 (1998).
\bibitem{gravitino} D. Delepine and U. Sarkar, hep-ph/9811479.
\bibitem{linde} G. Felder, L. Kofman, and A. Linde, hep-ph/9812289.
\bibitem{22} E. Ma, hep-ph/9807386 (Phys. Lett. {\bf B}, in press).
\end{thebibliography}

\newpage
\begin{table}
\begin{center}
\begin{tabular}{|c|c|c|c|}
\hline
superfield & gauge content & $Z_2$ & $Z'_2$ \\
\hline
$(N_1^0,E^-)$ & (1,2,--1/2) & -- & -- \\
\hline
$N_2^0$ & (1,1,0) & -- & -- \\
$E^+$ & (1,1,1) & -- & -- \\
\hline
$(h_3^0,h_3^-)$ & (1,2,--1/2) & + & -- \\
$(h_4^+,h_4^0)$ & (1,2,1/2) & + & -- \\
\hline
$\chi_1^+$ & (1,1,1) & + & -- \\
$\chi_2^-$ & (1,1,--1) & + & -- \\
\hline
\end{tabular}
\caption{New superfields added to the MSSM to obtain radiative neutrino 
masses.}
\end{center}
\end{table}
\begin{center}
\begin{picture}(360,200)(0,0)
\ArrowLine(0,0)(60,0)
\Text(30,-8)[c]{$\nu_i$}
\ArrowLine(120,0)(60,0)
\Text(90,-8)[c]{$h_3^-$}
\ArrowLine(120,0)(180,0)
\Text(150,-8)[c]{$h_4^+$}
\ArrowLine(240,0)(180,0)
\Text(210,-8)[c]{$\chi_2^-$}
\ArrowLine(240,0)(300,0)
\Text(270,-8)[c]{$\chi_1^+$}
\ArrowLine(360,0)(300,0)
\Text(330,-8)[c]{$\nu_j$}
\DashArrowLine(180,-60)(180,0)6
\Text(180,-72)[c]{$\langle h_2^0 \rangle$}
\DashArrowArc(180,-60)(134,90,154)6
\Text(100,60)[c]{$E^+$}
\DashArrowArcn(180,-60)(134,90,26)6

\Text(265,60)[c]{$E^-$}
\DashArrowLine(180,134)(180,74)6
\Text(180,146)[c]{$\langle h_2^0 \rangle$}
\end{picture}

\vspace{2.0in}
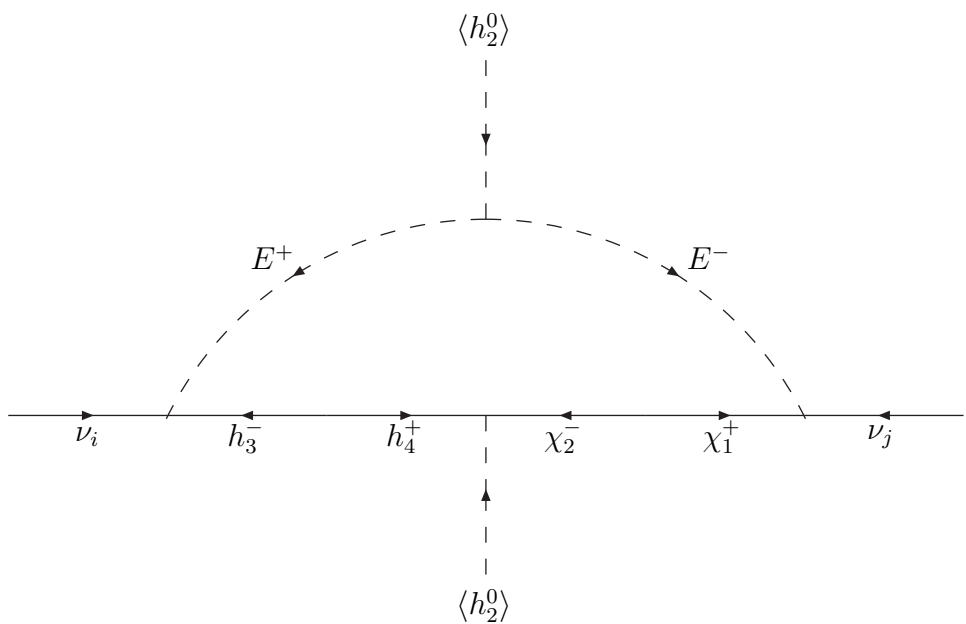
{\bf Fig.~1.} One-loop radiative generation of neutrino masses.
\end{center}
\end{document}